# Cross-section alignment of oblate grains

A.Lazarian [1,2] and M. Efroimsky [3]

astro-ph/9607137  25 Jul 1996

[1]Astronomy Department, University of Texas, Austin, TX 78712-1083

[2]Department of Astrophysical Sciences, Princeton University, Princeton NJ 08544 (present address)

[3]Department of Physics & Astronomy, Tufts University, Medford MA 02155

# ABSTRACT

This paper provides a quantitative account for a recently introduced mechanism of mechanical alignment of suprathermally rotating grains. These rapidly rotating grains are essentially not susceptible to random torques arising from gas-grain collisions, as the time-scales for such torques to have significant effect is orders of magnitude greater than the mean time between crossovers. Such grains can be aligned either by gaseous torques during short periods of crossovers or/and by the difference in the rate at which atoms arrive at grain surface arising from difference in orientation of grain in respect to the supersonic flow. The latter process, that we call cross-section alignment, is the subject of our present paper. We derive expressions for the measure of cross-section alignment for oblate grains, and study how this measure depends upon the angle between the interstellar magnetic field & the gaseous flow, and upon the grain shape.

*Subject headings:* magnetic fields – polarization – dust extinction

## 1. Introduction

In 1949, Hiltner and Hall (see Hall 1949) discovered that the reason for the starlight being polarized was orientation of grains of the interstellar dust. More precisely, they showed that the polarization was caused by linear dichroism (i.e., by the differential extinction of electromagnetic waves of different polarizations), which means that grains in the interstellar medium are aligned.

What makes dust grains (or, at least, a substantial share of them) prefer this or that orientation in space? The question is one of considerable nicety and difficulty (see discussion in Goodman et al 1995). To search for the possible mechanisms, one should recall several basic properties of the interstellar matter (hence, ISM). First, there exist enough evidence of paramagnetic nature of a substantial portion of grains (see Whittet 1992). This makes them subject to both paramagnetic relaxation [4] and to the Barnett effect, which produces magnetic moments of rotating grains. These momenta will then interact with the ubiquitous magnetic fields frozen into the interstellar plasma. The interaction will entail precession of the moments around magnetic lines. Note that the magnetic moment of a grain is always parallel to its vector of angular velocity, which is, roughly speaking, perpendicular to the longest dimension of the grain[5]

Strictly speaking, this is valid only for the suprathermal grains, i.e., for those whose rotational energy much exceeds that of the Brownian motion. It is these grains that are to be of our main concern in this article. We shall address them since a substantial part of grains in diffuse clouds should rotate suprathermally (Purcell 1979).

There may exist several spin-up mechanisms (Purcell 1975, 1979), and the catalytic one seems to be the most efficient. The essence of this mechanism is as follows. Normally, two separate atoms of hydrogen cannot form a molecule because it would have excessive energy and, thus, be unstable. The molecule may, however, be successfully produced on the surfaces that can trap one hydrogen atom in a potential well[6]; the second atom will react with it, while the excessive energy will be spent for getting the $H_2$ molecule out of the pit, and for acquiring an escape velocity. There may be many such potential wells on the surface of each grain (see Tielens & Allamandola 1987, Buch & Zhang 1991). Ejection of the nascent molecule will then entail mechanical back reaction, depositing angular momentum with the grains and, thus, making them subject to the suprathermal rotation.

---

[4]Being concerned with mechanical alignment, we do not discuss here paramagnetic relaxation, which has its own pros and cons (see Davis & Greenstein 1951, Jones & Spitzer 1968, Purcell 1969, 1975, 1979, Purcell & Spitzer 1971, Mathis 1986, Roberge, DeGraff & Flaherty 1993, Lazarian 1995 a,b,c).

[5]In presence of internal dissipation, a body tends to rotate around the axis of maximal moment of inertia. This occurs since the angular momentum is preserved and, for a fixed angular momentum, the body will have minimal kinetic energy of rotation, spinning around the axis of maximal inertia. As for the inner dissipation, it originates in the following way. The angular momentum $\mathbf{J}$ of the grain is a product of the inertia tensor by the angular velocity $\omega$. Generally, $\omega$ is varying in the coordinate system associated with the grain, and so does the magnetic moment $\mu$. As a result, magnetization varies in the grain coordinate system and this causes dissipation.

The second fact about the interstellar environment, that we wish to recall, is that the ISM is far from being quiescent. On the contrary, various flows are present. Among them are stellar winds, collimated outflows from young stellar objects and from accretion discs. Moreover, the entire ISM is dominated by supersonic MHD turbulence (see Aron & Max 1975, Myers 1985, 1987). All these motions of plasma, up to certain reservations, drag the dust grains, though with a lag caused by the grain inertia. The lag gives birth to the motion of the dust relative to the surrounding gas which, therefore, bombards the grains. Being charged grains also spiral about magnetic field (the Larmor frequency is $\omega_L$) lines and this also causes a relative grain-gas motion. A resonance acceleration of grains takes place when the Alfvénic frequency becomes equal to $\omega_L$. As the whole spectrum of frequencies starting from a cut-off frequency $\omega_c$ is present the interstellar medium it is possible to show, that the acceleration to large relative velocities must be present for sufficiently large grains for which $\omega_L < \omega_c$. The relevant detailed study of the dynamics of the charged grains is given in Lazarian & Draine (1996, in preparation).

All in all, if we disregard paramagnetic relaxation, which is marginal for ordinary (i.e., without superparamagnetic inclusions) paramagnetic grains in typical ISM conditions, the motion of a grain is determined, on the one hand, by its precession around the magnetic field line and, on the other hand, by the impacts produced by the flow of the surrounding gas.

The fact that a supersonic flow is capable of providing mechanical alignment was discovered by Gold (Gold 1951, 1952). Due to the presence of the precession, the process of Gold alignment becomes more complicated, as explained by by Dolginov & Mytrophanov (1976), Lazarian (1994) and Roberge, Hanany & Messinger (1995). Both paramagnetic and mechanical mechanisms were discussed by the above mentioned authors for thermally-rotating grains, i.e., for those whose angular momenta are predominantly determined by the Brownian motion. The paramagnetic mechanism was generalized by Purcell (1979) for the case of suprathermal motions. A similar generalization of the Gold mechanism was carried out by Lazarian (1995 d, f). Gold, in his pioneering papers, started out with the obvious fact that a flow interacting with a grain produces an angular momentum perpendicular to the flow. In case it exceeds the initial angular momentum (associated with the Brownian motion), the grain becomes aligned[7]. This consideration is, however, inapplicable to grains whose initial angular momenta are suprathermal (i.e. much exceed $kT_{gas}$).

---

[6]Such wells (the so-called active sites) may emerge on a grain surface due to, for example, inclusions or defects.

Nonetheless, there exist at least two other, also mechanical, processes that align suprathermal dust. The first of them, proposed by Lazarian (1995 d), was based on the circumstance that even a suprathermal grain becomes, for a short while, susceptible to the gaseous bombardment. This happens at the moment of a crossover, i.e., when the angular momentum is minimal.[8] The second process of mechanical suprathermal-grain alignment arises from the difference of cross-section of the grain-gas interaction (Lazarian 1995 d,f). The essence of this mechanism is as follows. The time between two crossovers depends on the rate of gas-grain collisions that is proportional to the gas-grain cross-section. Therefore, a grain will tend to spend more time in a position of the minimal cross-section. It is this effect that we are going to dwell upon in our article. One may inquire as to which of the two mentioned mechanisms will be of leading order in the realistic physical situations. As explained in Lazarian (1995d), the answer depends upon the gas velocity. For slow motions ($\leq 2$ km/s), the random torques caused by hydrogen molecule creation exceed those entailed by the gaseous bombardment and, therefore, only marginal alignment may arise during the crossovers. Thus, only the second, cross-section mechanism is essential. For high-velocity flows, both the crossover and the cross section mechanisms are relevant: their contributions to the total effect are of about the same order and, what is especially remarkable, they coerce the grain to align in the same direction. It happens as the crossover mechanism also tends to minimize the grain cross-section in respect to a gaseous flow (Lazarian 1995 d). All these estimates are, in fact, valid for oblate grains only. For the prolate ones, the cross-section mechanism is weak (Lazarian 1995 d). However, there is evidence (Draine and Lee 1984, Hildebrand 1988, Hildebrand & Dragovan 1995) that at least at some regions it is oblate grains that dominate polarization.

---

[7]It is easy to show, that for the mechanical alignment in general and for the Gold alignment in particular supersonic motions of grains are necessary.

[8]The crossover happen, for example, due to grain resurfacing (Spitzer & McGlynn 1979) caused by accretion: when a monolayer is accreted, active sites emerge on new places. This may cause a reversal of the torque direction.

Now, we are going to present a detailed calculation of the measure of alignment for oblate grains. In the next section, we shall state the topic in the mathematical setting. In section 3, the solution to the problem is presented in a general form, while in section 4 some particular cases are addressed. The final section, 5, is devoted to a brief discussion of the obtained results.

## 2. Statement of the problem

Our principal goal is to find the measure of grain orientation arising from the cross-section mechanism. According to Greenberg (1968), the amount of alignment can be expressed by the so-called Rayleigh reduction factor, $\sigma$, that relates the statistical properties of the grain ensemble to the properties of the observed polarized radiation.

Let $z$ be the direction of propagation of the observed radiation, and $x$, $y$ be the directions of the maximum and minimum electric wave amplitudes $\mathbf{E}$. Let $s$ be the axis of rotation of a spinning grain. Mean absorption cross-sections of (partially aligned, spinning) grains for $\mathbf{E} \| x$ and $\mathbf{E} \| y$ will be denoted as $C_x$ and $C_y$, respectfully, while those for $\mathbf{E} \| \mathbf{s}$ and $\mathbf{E} \perp \mathbf{s}$ will be named as $C(\mathbf{E} \| \mathbf{s})$ and $C(\mathbf{E} \perp \mathbf{s})$. Then, according to Hildebrand (1988), Appendix,

$$|C_x - C_y| = (C(\mathbf{E} \| \mathbf{s}) - C(\mathbf{E} \perp \mathbf{s}))\sigma \cos^2 \gamma, \tag{1}$$

where $\gamma$ is the angle at which the direction, $\mathbf{B}$, of the magnetic field is inclined to the line of sight. Then, for spheroidal grains,

$$\sigma = \frac{3}{2} \left( \langle \cos^2 \varphi \rangle - \frac{1}{3} \right), \tag{2}$$

where the angular brackets stand for an ensemble average, and $\phi$ is the angle between the magnetic field and the symmetry axis of a grain. As we already mentioned (see the comment on page 1), the direction of the magnetic moment will be parallel to the axis of symmetry, for oblate spheroidal grains, and will be perpendicular to it for prolate ones. So for oblate grains the angle $\varphi$ will be the one between the magnetic moment and the magnetic field.

The factor $\sigma$ can indeed be considered as a measure of alignment. In absence of alignment, ensemble average of $\cos^2 \varphi$ in Eq.(2) become equivalent to averaging over the solid angle: $\langle \cos^2 \varphi \rangle = 1/3$, and thence $\sigma = 0$. This gets along with $C_x$ being equal to $C_y$. For perfect alignment, $\langle \cos^2 \varphi \rangle = 1$ and, thus, $\sigma$ is maximal, i.e., unity.

Thus, our goal now is to calculate $\sigma$. As explained above, there are indications of the polarization effect being mainly produced by oblate grains. So we shall model the grain by an oblate spheroid. To compute the average of $\cos^2 \varphi$ over an ensemble, one needs to introduce a distribution function $\rho$ and to execute a normalization

$$\langle \cos^2 \varphi \rangle = \frac{\int \cos^2 \varphi \rho \mathrm{d}\Omega}{\int \rho \mathrm{d}\Omega}, \tag{3}$$

which is to be done in the next section.

## 3. Calculation of the Rayleigh reduction factor

For a grain taken at random, its mean time back to crossover, $t_x$, will be of order of the mean time between two sequent crossovers, though slightly different. According to Purcell (1979) the formula,

$$t_x \approx 1.3(t_L + 0.6t_d) \tag{4}$$

provides a good approximation for $0.1 < t_L/t_d < 10$, $t_L$ being a typical time of existence of active sites, and $t_d$ being the rotational damping time (the period during which the gaseous friction would stop rotation of a free rotating grain).

We do not specify here, whether the life-time of active sites is the time necessary for accreting a monolayer of refractory material (see Spitzer & McGlynn 1979) or the time of poisoning of the active sites (see Lazarian 1995a,f) as our conclusions do not depend on the particular process limiting the life of Purcell's rockets. Indeed, what is important for us is that, for a chosen grain, $t_L$ is the time of accreting of a particular number of heavy atoms. In its turn $t_d$ is proportional to the time of grain colliding with the mass of gas equal to the grain mass.

To understand the essence of the mechanism proposed, consider at first a toy model, namely, assume that the axis of suprathermally rotating oblate grain can have only two positions, namely perpendicular ($\perp$) and parallel ($\parallel$) to the axis of a gaseous flux. In this model, grain axis stays for the time $t_x$ in one position and then undergoes the crossover and has equal chances to get either the same or the perpendicular alignment[9]. In the absence of the gaseous flux time scales $t_{x\parallel}$ and $t_{x\perp}$ are the same and, naturally, there is no preferential position. However, it is easy to see, that in presence of the gaseous supersonic flux, $t_{x\parallel}$ and $t_{x\perp}$ are inversely proportional to gas-grain cross sections $S_{n\parallel}$ and $S_{n\perp}$ respectively.

---

[9]A complete disorientation is a good model for sufficiently large grains as according to eq. (A11) in Lazarian (1995a). For small grains the time of complete disorientation can be of the order of several $t_x$. This, however, does not change anything in our conclusions.

As the time of the crossover is negligible compared with any of $t_x$, we conclude that the time averaged probability of finding our grain in positions $\perp$ or $\parallel$ is inversely proportional, respectively, to $S_{n\perp}$ and $S_{n\parallel}$. According to the ergodic hypothesis this probability coincides with the ensemble averaged one.

Returning from the toy model to real grains we observe, that in this case both $t_L$ and $t_d$ are inversely proportional to the rate at which the atoms arrive at the grain surface and this rate is proportional to the effective cross-section of the grain, which, in its turn, depends upon the angles $\Phi$, $\varphi$ and $\Psi$, as on Fig. 1. On this figure, $\mathbf{H}$ denotes the magnetic field, $\mathbf{J}$ stands for the angular momentum, $\mathbf{u}$ is the velocity of the gas; angle $\Phi$, between $\mathbf{H}$ and, $\mathbf{u}$, varies from 0 to $\pi$; angle $\varphi$, between $\mathbf{H}$ and $\mathbf{J}$, varies from 0 to $\pi$, while the angle $\Psi$, between the projections of $\mathbf{J}$ and $u$ on the plane orthogonal to $\mathbf{H}$, may vary from 0 to $2\pi$. In the course of precession, the vector $\mathbf{J}$ will sweep a cone, of angle $\varphi$, around the magnetic line (Fig. 1). As proven in numerous papers ( Dolginov & Mitrofanov 1976; Purcell 1979, Roberge et al. 1993), the period of the precession is many orders less than the time $t_x$ that has elapsed since the angular velocity of a grain spinning around its own axis last crossed through zero. Since the grain preserves the angle $\varphi$ during the time between crossovers, and since, by definition, the probability to find a grain at a position given by angle $\varphi$ is proportional to the time the grain spends in this position, one can state that this probability is proportional to the mean time between two consecutive crossovers first of which corresponded to $\varphi$.

Damping of spinning is caused both by inelastic collisions and by grain resurfacing that changes the positions of active sites. Suppose that on average damping takes place after a grain undergoes $N_1$ inelastic collisions, while for a full resurfacing $N_2$ collisions are sufficient. Let $N_m = max(N_1, N_2)$; then a crossover will happen after approximately $N_m$ collisions. The frequency of collisions is proportional to the effective cross-section area corresponding to a certain position of the grain. Thence, the mean time between 2 consecutive collisions is inversely proportional to this effective cross-section. This cross-section varies as a grain precesses. Consequently the time $t$ for $N_m$ collisions to occur[10] depends on $\varphi$:

$$N_m = \int_0^t nuS(t)\mathrm{d}t, \tag{5}$$

where $S(t)$ is cross section of gas-grain interaction, which varies with time. For $t$ much greater than the period of precession $T$ Eq(5) gives

$$N_m = \frac{t}{T}\int_0^T nuS(t)\mathrm{d}t \tag{6}$$

---

[10]We are happy to acknowledge here, that the elegant derivation of the time between crossovers below was pointed to us by Bruce Draine.

Inversion of the equation above provides:

$$t = \frac{N_m}{\int_0^T nu S(t) dt} = \frac{N_m}{nu T \frac{1}{2\pi} \int_0^{2\pi} S(\Psi) d\Psi},$$ (7)

where it was taken into account, that the angular velocity of precession $\omega$ is equal to $\frac{d\Psi}{dt}$. The probability to find a grain with a given $\Phi$ is proportional to time $t$ that is stays in this position. In fact, Eq.(7) provides the non-normalized distribution function. To normalize it, one should integrate (7) also over $\varphi$, thereby getting for any fixed $\Phi$ the normalization $C$. Finally, one will get, for the Raleigh reduction factor:

$$\sigma = \frac{3}{2} \frac{I}{C} - \frac{1}{2},$$ (8)

where

$$I = \int_0^\pi d\varphi \sin\varphi \cos^2\varphi \frac{1}{\int_0^{2\pi} d\Psi S(\Phi, \varphi, \Psi)}$$ (9)

and

$$C = \int_0^\pi d\varphi \sin\varphi \frac{1}{\int_0^{2\pi} d\Psi S(\Phi, \varphi, \Psi)}$$ (10)

($\sin\varphi$ appears since the integrals in Eqs.(10) and (9) are to be taken over the solid angle).

To compute the effective cross-section $S(\Phi, \varphi, \Psi)$, we approximate the grain by an oblate spheroid of a long semi-axis $a$, and a short semi-axis $b$ (Fig. 2). In Fig. 3, $\mathbf{e}$ is the unit vector in the direction of the gas flow, while $\mathbf{n}$ is the normal to the appropriate cross-section of the ellipsoid[11](note that this is a regular cross-section, not the effective one which is to be perpendicular to the gaseous flow). Let $\nu$ be the angle between the cross-section and the axis $y$. For $x = 0$, $y = \rho \cos\nu$, $z = \rho \sin\nu$, and $\frac{y^2}{a^2} + \frac{z^2}{b^2} = 1$, so one can easily express $\rho$ via $a$, $b$ and $\nu$:

$$\rho = \left( \frac{\cos^2\nu}{a^2} + \frac{\sin^2\nu}{b^2} \right)^{-1/2}.$$ (11)

Thus, the area of the figure darkened on Fig. 2 (the regular cross-section) is

$$\Sigma = \pi a \rho = \frac{\pi a}{\sqrt{\frac{\cos^2\nu}{a^2} + \frac{\sin^2\nu}{b^2}}} = \frac{\pi a}{\sqrt{\cos^2\nu + (a/b)^2 \sin^2\nu}}$$ (12)

Now, let $\alpha$ denote the angle between the direction of the gaseous flow and the short axis of the spheroid. Then one has, for the unit vector $\mathbf{e}$: $e_y = \sin\alpha$, $e_z = \cos\alpha$. Similarly, for $\mathbf{n}$:

---

[11]The circumference of this elliptical cross section corresponds to the points at which the gaseous flow is tangential to the grain surface.

$n_y = \sin \nu$, $n_z = \cos \nu$. The projection of the area (the regular cross section) to the direction will be the effective cross-section we are looking for:

$$S = \Sigma |\mathbf{n}\mathbf{e}| = \pi a^2 \frac{\sin \nu \sin \alpha + \cos \nu \cos \alpha}{\sqrt{\cos^2 \nu + (a/b)^2 \sin^2 \nu}} \tag{13}$$

An observer looking at the grain along the vector $\mathbf{e}$ will notice an elliptic feature of the effective area $S$. To get rid of the angle $\nu$ in the latter expression, we would notice that, for $x = 0$,

$$y = a\sqrt{1 - z^2/b^2} \tag{14}$$

and

$$\tan \alpha = -(\partial y/\partial z)_{x=0} = (a/b)^2 \tan \nu \tag{15}$$

Plugging this into (13) we obtain an expression for $S$, in terms of $a$, $b$, and $\alpha$ only:

$$S = \pi a^2 \cos \alpha \frac{(b/a)^2 \tan^2 \alpha + 1}{\sqrt{1 + (b/a)^2 \tan^2 \alpha}} = \pi a^2 \sqrt{\cos^2 \alpha (1 - (b/a)^2) + (b/a)^2} \tag{16}$$

Substitution of this in Eqs. (9) and (10) yields the following expressions for the quantities entering formula (8) for the Rayleigh reduction factor:

$$I = \int_0^\pi \mathrm{d}\varphi \sin \varphi \cos^2 \varphi \left[ \int_0^{2\pi} \mathrm{d}\Psi \sqrt{\cos^2 \alpha (1 - (b/a)^2) + (b/a)^2} \right]^{-1} \tag{17}$$

$$C = \int_0^\pi \mathrm{d}\varphi \sin \varphi \left[ \int_0^{2\pi} \mathrm{d}\Psi \sqrt{\cos^2 \alpha (1 - (b/a)^2) + (b/a)^2} \right]^{-1} \tag{18}$$

Since the ratio of I and C enter the expression for the Rayleigh reduction factor (see Eq. (8)) we omitted in both Eqs. (17) (18) the multipliers $1/(\pi a^2)$. It follows from Fig.1 that

$$\cos \alpha = \cos \Phi \cos \varphi + \sin \Phi \sin \varphi \cos \Psi. \tag{19}$$

In section 4 we present analytical results obtained for particular physically important cases and numerical computations for other cases.

## 4. Results

### 4.1. Analytics for particular cases

Of practical value is the case of the gaseous flow being parallel to the magnetic field. This corresponds to the situation when the dust is driven by radiation pressure (see Spitzer 1978). Bearing charge, these grains are coerced to follow magnetic field lines. In the reference system associated with the moving dust, the gas will move in an opposite direction, i.e., parallel to the magnetic field. (Such motions are typical for bright-star vicinities.) In this case, the integrals (9) and (10) can be taken analytically (see Appendix A):

$$\frac{I_{\Phi=0}}{C_{\Phi=0}} = \frac{1}{2}\frac{(b/a)^2}{1-(b/a)^2}\frac{(a/b)^2\sqrt{1-(b/a)^2}-\ln(a/b+\sqrt{(a/b)^2-1})}{\ln(a/b+\sqrt{(a/b)^2-1})} \tag{20}$$

The next case of practical importance is that of Alfvenic waves in the interstellar plasma. Their interaction with dust is to be discussed elsewhere; here we would just mention that according to observations these waves result in supersonic motion of interstellar gas (Aron and Max 1975; Myers 1987). The relative direction of grain-gas motion corresponds to $\Phi = \pi/2$ ( Lazarian 1994). (A similar situation is inherent to the ambipolar diffusion mechanism (Roberge & Hanany 1990, Roberge, Hanany & Messinger 1995)). For hypersonic flows, it is sufficient to calculate $I$ for exactly $\Phi = \pi/2$, and the appropriate asymptotics is derived in Appendix B, in assumption of the parameter

$$F = 1 - (b/a)^2 \tag{21}$$

being small, i.e., $b/a$ being close to unity. Here follows the asymptotic:

$$\left(\frac{I}{C}\right)_{\Phi=\frac{\pi}{2},\frac{1}{2}<\frac{b}{a}<1} \approx \frac{1}{3} + \frac{1}{45}F + \frac{2}{135}F^2. \tag{22}$$

We show below, that this approximation is sufficiently accurate for $b/a > 2/3$.

Finally, we wish to address the extreme case of dust grains modelled by discs. For an absolutely flat disc (b = 0), calculations for $\Phi = \pi/2$ provide $\sigma = 0.25$. This is the maximal alignment for $\Phi = \pi/2$.

## 4.2. Other cases

Our analytical studies gave us the grasp of the mechanism under study. To get more details we will use direct numerical computation of the integrals involved. For instance, the measures of alignment for grains with a number of different axis ratios are shown in Fig. 4 for $\Phi$ from 0 to $\pi/2$. Naturally, the difference in cross sections is increasing with the decrease of the $b/a$ ratio and therefore the measure of alignment should increase as the result.

Fig. 4 proves this tendency. For grains with the axis ratio more than 2/3 the alignment is marginal and therefore the influence of the mechanism discussed on the alignment of sufficiently spherical grains is negligible. In Lazarian (1994, 1995d) the anisotropy of gaseous bombardment was characterized by the parameter $s$, which is proportional to $\langle \cos^2 \Phi - 1/3 \rangle$. It was shown there, that for both the uniform bombardment corresponding to $\langle \cos^2 \Phi \rangle = 1/3$ and the bombardment by an ordered gaseous flux at $\Phi = \arccos 1/\sqrt{3}$ produce no alignment of grains. We see that the same result is valid for cross-section alignment, namely the alignment vanishes for $\Phi = \arccos 1/\sqrt{3}$.

Alignment of grains subjected to Alfvénic waves and ambipolar diffusion is shown in Fig. 5. It is easy to see that for grains with the axis ratio greater than 0.1 the measure of cross section alignment does not exceed 20%.

Grains under the radiation pressure follow magnetic field lines and the measure of such an alignment is shown in Fig. 6. The efficiency of alignment also decreases with the increase of the axis ratio, but for very small axis ratios, i.e. $b/a \ll 0.1$ the alignment is more efficient as compared with the previous case.

## 5.   Discussion

In the article thus far, we have considered alignment of suprathermally rotating oblate grains. The alignment is produced in respect to interstellar magnetic field as a grain rapidly precesses about this direction. This precession is taking place due to the fact that grains are subject to the Barnett effect, i.e., they acquire a magnetic moments as a result of rapid spin-up. The alignment itself arises from grain interaction with gaseous flow.

We have shown that the measure of alignment (the so-called Rayleigh reduction factor) is larger for more oblate grains, reaching its maximum for infinitesimally flat disc grains (we call them flakes). For example, when $\Phi = 0$ (gas flow moves along the magnetic line), the said measure [12] reaches $-0.5$, while for $\Phi = \pi/2$ (gas flow perpendicular to the field), it is 0.25.

The most striking feature of the discussed mechanism of alignment is that it (unlike the Gold alignment) does not depend on the angular momentum that deposited with a grain by gaseous flow. It is only the rate, at which atoms arrive at the grain, that matters.

---

[12]The negative sign of the measure corresponds to preferential alignment of grain long axis along to magnetic field, while positive measures correspond to the preferential alignment perpendicular to the field.

As this rate depends on the grain cross section relative to the gaseous flow, this mechanism is called cross-section alignment. To find the direction of the alignment, the rule of thumb is that a grain tends to minimize its cross section in respect to the gaseous flow.

In the vicinity of bright sources grains subjected to the radiation pressure follow magnetic field lines. As a result, the cross section mechanism coerce grains to align along magnetic field lines. The competing process arises from ubiquitous Alfvénic waves which tend to align grains perpendicular to magnetic field lines. In the absence of other alignment mechanisms, the interplay between these two effects would determine the polarization pattern seen in the ISM. We believe that in reality in different regions of the ISM different processes dominate, and one can distinguish these regions only on the basis of comparing quantitative theoretical predictions with observations. Such a work for a particular region is being done in (Lazarian, Goodman & Myers 1996 in preparation).

Above we discussed only a very simple model. Indeed, it has been assumed, that the flow is hypersonic (so we may disregard stochastic motions of individual atoms), and that the direction of the flow does not change in the course of alignment. This is not always true. For instance, large-amplitude Alfvénic waves coerce grains to move in such a way that the angle between magnetic field and the direction of grain motion is permanently changing. In another example, the velocities induced by ambipolar diffusion are usually only mildly supersonic and therefore one must take into account the rms velocities of atoms. Fortunately, our model can be easily generalized to account for these complications. The way for this generalization was discussed in Lazarian (1995d). Without going into details, we can exemplify the suggested procedure for the case of rms velocities. In this case it is possible to express the angle $\Phi$ in the following way:

$$\Phi = \arctan \sqrt{\frac{(U_x + v_x)^2 + (U_y + v_y)^2}{(U_z + v_z)^2}}, \tag{23}$$

where $v_i$ $i = x, y, z$ are the rms velocity components, while $U_e$ are the components of the flow velocity. To obtain the corresponding measures of alignment, one needs to integrate over the Maxwellian distribution. Obviously enough, only supersonic flows can provide substantial alignment; otherwise, if $|U| < \sqrt{v^2}$, averaging over $\Phi$ makes the alignment negligible.

In the real world, other alignment processes will also act on the suprathermally rotating grains, the crossover (Lazarian 1995d) and paramagnetic (Purcell 1979) mechanisms being just two examples. If one knows the measures of alignment for individual processes, i.e. $\sigma^{(1)}$ and $\sigma^{(2)}$, it is possible to find the resulting measure of alignment $\sigma^{(\sum)}$ when the two processes act simultaneously (Lazarian 1995d):

$$\sigma^{(\Sigma)} \approx \frac{\sigma^{(1)} + \sigma^{(1)}\sigma^{(2)} + \sigma^{(2)}}{1 + 2\sigma^{(1)}\sigma^{(2)}}. \tag{24}$$

Naturally, the above rule can easily be generalized on any number of processes acting simultaneously.

Speaking about the mechanical alignment only, one should bear in mind that, for mildly supersonic velocities of grain-gas relative motion, the rms torques associated with $H_2$ formation may dominate rms torques arising from atomic impacts during crossovers, i.e., $\gamma \langle (\delta J_{H2})^2 \rangle > \langle (\delta J_{collision})^2 \rangle$, where $\gamma^{-1}$ is the ratio of the atoms striking the grain to that included in the $H_2$ molecules formed over grain surface. In this case, the alignment measure corresponding to crossover mechanism should be diminished by a factor $x/(1 + x)$ where $x = \langle (\delta J_{collision})^2 \rangle / [\gamma \langle (\delta J_{H2})^2 \rangle]$. If $x \ll 1$, the contribution from the crossover alignment becomes negligible. Similarly, for ordinary paramagnetic grains in typical ISM conditions, the alignment is negligible provided that the spin-up is short-lived (Spitzer & McGlynn 1979). All in all, there exists a range of conditions where the cross-section alignment is the dominant one.

This work was initiated by Bruce Draine's comments, but would not be possible if not for encouragement by Ethan Vishniac. We also indebted to Bruce Draine for pointing to us weak points of the original version of the paper and to the referee Wayne Roberge for a number of helpful comments. A.L. gratefully acknowledges the support from NASA grants NAG5 2773 at the University of Texas at Austin and NAG5 2858 at Princeton University. A.L. thanks Alyssa Goodman, Walt Duley, Jim Hough, Phil Myers and Lyman Spitzer for illuminating discussions.

## Appendix

## A. Grains streaming along magnetic field lines

Streaming of grains corresponds to $\Phi = 0$. The general expression in this case can be simplified and the integration over $\Psi$ is equivalent to multiplication by $2\pi$. The normalization in this case is equal to

$$C = 2 \int_0^{\pi/2} \frac{\sin\varphi \, d\varphi}{2\pi \sqrt{F \cos^2 \varphi + E}}, \tag{A1}$$

where

$$F = 1 - (b/a)^2 \tag{A2}$$

and

$$E = (b/a)^2. \tag{A3}$$

Integral (A1) can be easily calculated by substituting $z = \sin(\varphi/2)$

$$C = \frac{1}{\pi\sqrt{1 - (b/a)^2}} \ln\left(\frac{\sqrt{1 - (b/a)^2} + 1}{(b/a)}\right). \tag{A4}$$

Similarly the integral $I$ provides:

$$I = 2\int_0^{\pi/2} \frac{\cos^2\varphi\sin\varphi\,\mathrm{d}\varphi}{2\pi\sqrt{F\cos^2\varphi + E}} = \frac{E}{2\pi F^{3/2}}\left\{\frac{\sqrt{F}}{E} - \ln\frac{\sqrt{F}+1}{\sqrt{E}}\right\}. \tag{A5}$$

All in all the ratio $I/C$ is equal to

$$I/C = \frac{1}{2}\frac{(b/a)^2}{1-(b/a)^2}\frac{(a/b)^2\sqrt{1-(b/a)^2} - \ln(a/b + \sqrt{(a/b)^2 - 1})}{\ln(a/b + \sqrt{(a/b)^2 - 1})} \tag{A6}$$

and this ratio approaches zero for evanescent $b/a$.

## B. Grain drift perpendicular to magnetic lines

This case corresponds to $\Phi = \pi/2$. The integral over $\Psi$ can be rewritten using the notations of the previous section:

$$J = 4\int_0^{\pi/2} \mathrm{d}\Psi\sqrt{\sin^2\varphi\cos^2\varphi F + E}. \tag{B7}$$

This integral, according to the formula [2.614] in Gradshteyn & Ryzkik (1965), reads:

$$J = 4\sqrt{E + F\sin^2\varphi}\,\mathcal{E}\left(\frac{\sqrt{F}\sin\varphi}{\sqrt{F\sin^2\varphi + E}}\right) \tag{B8}$$

where the complete elliptic integral E of the second kind can be expanded for $b/a \to 1$ in the following way:

$$\mathcal{E}(k) \approx \frac{\pi}{2}\left[1 - \frac{1}{4}k^2 - \frac{3}{64}k^4 - \frac{5}{256}k^6 - ...\right]. \tag{B9}$$

By means of the latter, the reciprocal former may be presented as

$$J^{-1} \approx \frac{1}{2\pi\sqrt{F}\sin\varphi}\left(k + \frac{1}{4}k^3 + \frac{7}{64}k^5 + \frac{15}{256}k^7 + ...\right) \tag{B10}$$

where $k = \sqrt{F}\sin\varphi/\sqrt{F\sin^2\varphi + E}$. So the normalization will read

$$C \approx \frac{1}{4\pi}\left[4 + \frac{4}{3}F + \frac{22}{30}F^2 + \frac{19}{48}F^3\right]. \tag{B11}$$

Similarly, for the quantity $I$:

$$I \approx \frac{1}{\pi}\left[\frac{1}{3} + \frac{2}{15}F + \frac{1}{12}F^2 + \frac{155}{576}F^3\right]. \tag{B12}$$

Now we are well equipped to compute the ratio $I/C$:

$$\frac{I}{C} \approx \frac{1}{3} + \frac{1}{45}F + \frac{2}{135}F^2 + .... \tag{B13}$$

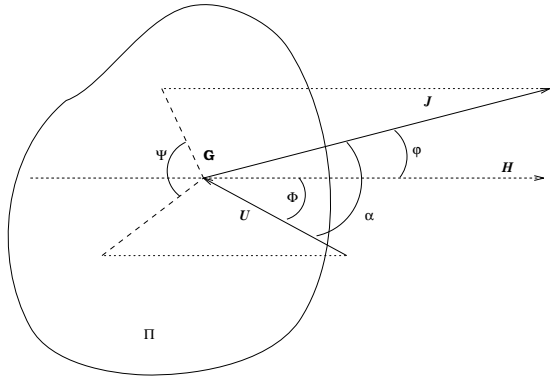

Fig. 1.— Grain at the position **G** is subjected to the flux in the direction given by the relative velocity **u** and its short axis is directed along **J**. The angle $\psi$ is measured in the plane $\Pi$ which is perpendicular to the direction of magnetic field **H**.

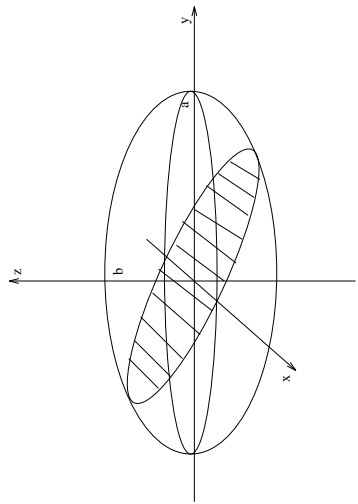

Fig. 2.— Cross section of a grain.

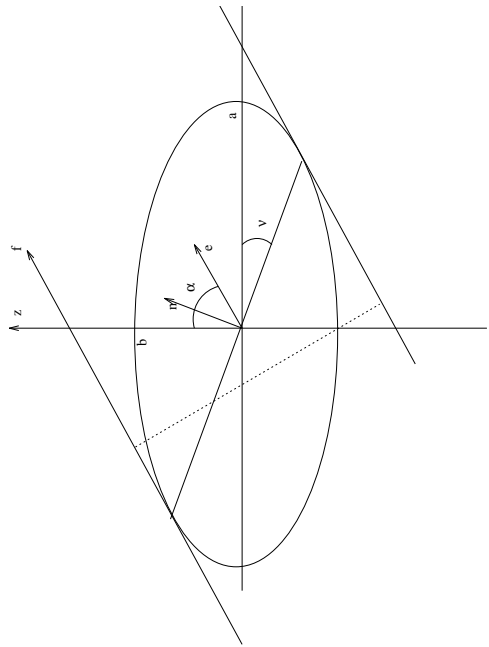

Fig. 3.— Cross section of a grain. The dotted line corresponds to the crossection $S$.

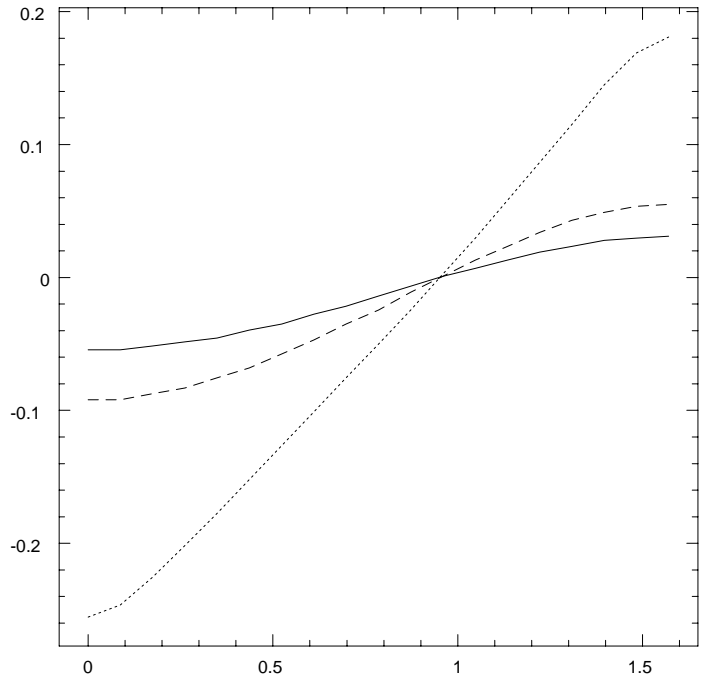

Fig. 4.— Measure of alignment $\sigma$ as a function of a $\Phi$, which is the angle between $\mathbf{H}$ and $\mathbf{u}$. *Phi* changes from 0 to $\pi/2$. When $\Phi = \arccos 1/\sqrt{3}$ all the curves pass through zero. The solid line corresponds to $b/a = 2/3$, short dashed line corresponds to $b/a = 0.2$, while the dotted line corresponds to $b/a = 0.1$.

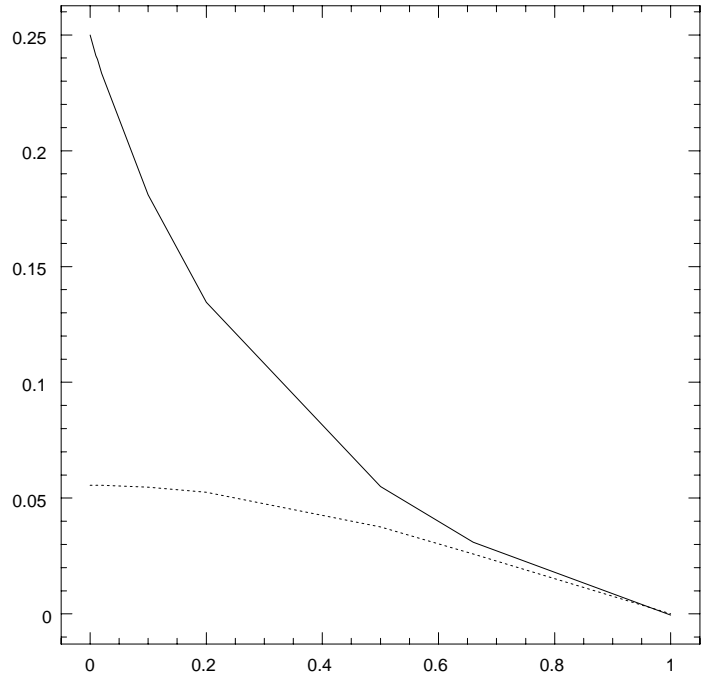

Fig. 5.— Measure of alignment of grains subjected to Alfvénic waves or ambipolar diffusion as a function of the axis ratio. This ratio changes from zero (maximal alignment) to one (no alignment at all). The dashed line corresponds to the asymptotic obtained for small F.

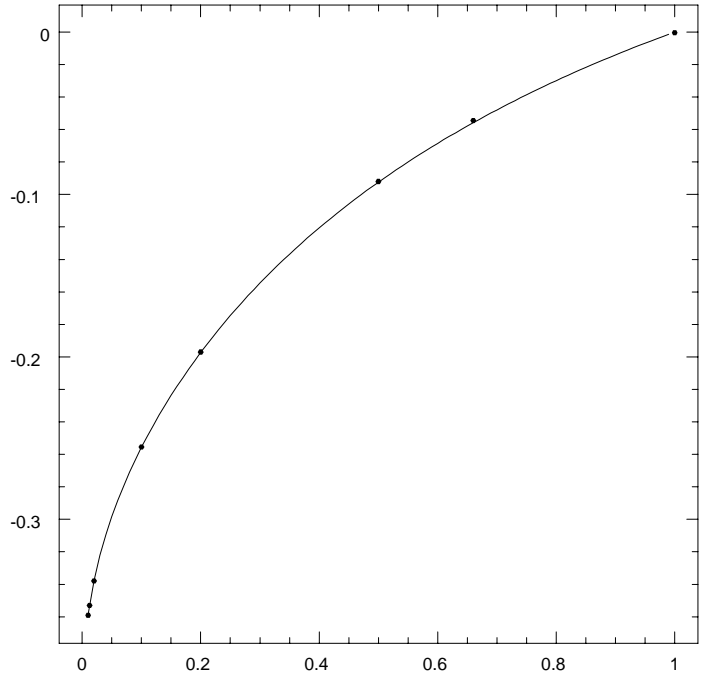

Fig. 6.— Measure of alignment of grains streaming along magnetic field lines as a function of the axis ratio. This ratio changes from zero (maximal alignment) to one (no alignment at all). The solid line corresponds to the analytical solution found, while the filled hexagons corresnpond to numerical computations.